\documentstyle[epsf]{mn}

\begin{document}

\title[037--B327 in M31]{The `remarkable' M31 globular cluster 037--B327 revisited}

\author[Barmby, Perrett, \& Bridges]{Pauline Barmby,$^1$ Kathryn M. Perrett,$^2$ Terry J. Bridges$^3$\\
$^1$Harvard-Smithsonian Center for Astrophysics, 60 Garden St., Cambridge, MA 02138\\
$^2$Department of Physics, Queen's University, Kingston, Ontario K7L 3N6, Canada\\
$^3$Anglo--Australian Observatory, Epping, NSW, 1710 Australia}

\maketitle

\begin{abstract}

The M31 globular cluster candidate 037--B327 has long been known to
be an extremely red, non-stellar object. The first published spectrum of
this object is used to confirm that it
is a globular cluster belonging to M31, with rather typical values of
$v_r=-338\pm 12{\rm~km~s}^{-1}$ and ${\rm [Fe/H]}=-1.07\pm 0.20$~dex. 
Using the spectroscopic metallicity to predict the intrinsic
colours, we derive a reddening value of $E(B-V)=1.30\pm0.04$,
in good agreement with the value obtained using reddening-free
parameters. The extinction-corrected magnitude of 037--B327 is $V_0=12.73$ 
(absolute magnitude $M_V=-11.74$),
which makes it the most luminous globular cluster in M31. 
We examine van den Bergh's (1968) argument regarding the
brightest and most-reddened globular cluster in in M31; we find 
that the brightest clusters are more heavily-reddened than average, but
this can be explained by selection effects rather than a different $R_V$ in M31.

\end{abstract}


\section{Introduction}

The study of globular clusters in M31 dates back to Hubble \shortcite{hub32}.
Since that pioneering work, over a thousand objects have been proposed
as possible M31 globular clusters (GCs), and over 200 have been 
confirmed as belonging to M31. The M31 GCS has long been one of the
standard extragalactic globular cluster systems used for comparison to
the Milky Way GCS, and comparisons of the two systems have been both 
fruitful and puzzling. 

Few individual M31 GC candidates have attracted as much interest as 
the object known as B327 (B for `Baade'), Bo037 (Bo for `Bologna'), 
or 037--B327 [in the nomenclature introduced by Huchra, Brodie \& Kent 
\shortcite{hbk91}].
This object (Figure~\ref{fig-finder}) was first identified 
as a globular cluster candidate by Baade, using plates taken in c.\ 1945.
A portion of Baade's M31 GC candidate list appears in Seyfert \& Nassau \shortcite{sn45},
but the coordinates of 037--B327 were first published by Vete\v{s}nik \shortcite{vet62a}.
Kron \& Mayall \shortcite{km60} first measured an extremely red colour for this object
and suggested that it must be highly reddened and extremely luminous.
Vete\v{s}nik \shortcite{vet62b} found that 037--B327 was the most highly-reddened
in his sample of 209 objects with $E(B-V)=1.28$. van~den~Bergh \shortcite{vdb68,vdb69} 
used these results to argue that the value of $R_V=A_V/E(B-V)$ in M31 was 2.5, 
instead of the Milky Way value of 3.0.
Sargent et al.\ \shortcite{sar77} rejected 037--B327 from their M31 GC candidate list
because of its stellar appearance, but Buonanno et al.\ \shortcite{buo82} `[confirmed]
its non-stellar nature' using measurements of its image size
on photographic plates.
Using low-resolution spectroscopy, Crampton et al.\ \shortcite{cra85} again found
037--B327 to be the most highly-reddened GC candidate in M31, with 
$E(B-V)=1.48$. Sharov \& Lyutyi \shortcite{sl89} concluded that it was the most
luminous GC in M31, while cautioning that its true nature 
was still unknown. Barmby et al.\ \shortcite{b00} echoed both of these results:
they measured $E(B-V)=1.38$ and $V_0=12.54$, corresponding
to a luminosity four times that of the brightest Milky Way GC.

\begin{figure}
\epsfxsize=8cm
\caption{Digitized Sky Survey image of $30\times30$\arcmin\ region of M31, centered
on coordinates 00:42:02.9 +41:14:38.0 (J2000). 037--B327 is the circled object, coordinates 
00:41:35.0 +41:14:54.5 (J2000). \label{fig-finder}} 
\end{figure}

In this paper, we confirm that 037--B327 is an M31 globular 
cluster. We discuss its photometric and spectroscopic properties 
and provide a new estimate of its reddening. Since van~den~Bergh's
arguments about the value of $R_V$ in M31 were motivated in part by the
properties of 037--B327, we revisit that argument using the properties
of this cluster and others reported in Barmby et al.\ \shortcite{b00}.

\section{Properties of 037--B327}

\begin{figure}
\epsfxsize=8cm
\epsfbox{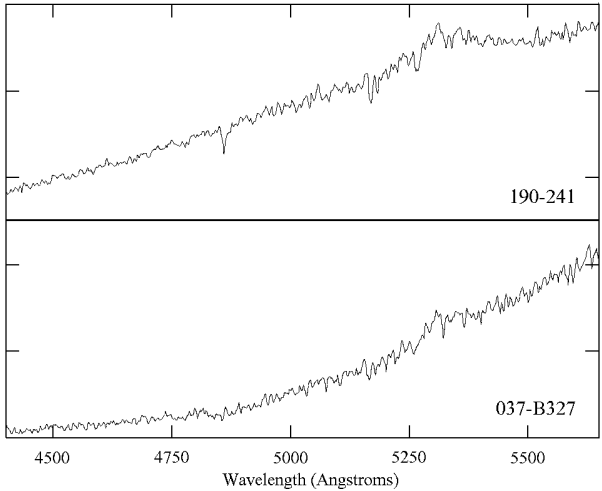}
\caption{Spectrum of 037--B327, with spectrum of 190--241
(obtained during the same integration;
${\rm [Fe/H]}=-1.03 \pm 0.09$) shown for comparison.
The ordinate scale of the graph is linear, in arbitrary units.
Note the depressed red end of the 037--B327 spectrum. \label{fig-spectrum}}
\end{figure}

As part of a larger project, Perrett et al.\ (2001, in preparation) obtained spectra of 
about 200 M31 globular clusters using the WYFFOS
Wide-Field Fibre Optic Spectrograph at the William Herschel 4.2~m
telescope\footnote{The William Herschel Telescope is operated on the
island of La Palma by the Isaac Newton Group in the Spanish
Observatorio del Roque de los Muchachos of the Instituto de
Astrofisica de Canarias.} in November 1996.  The R1200R grating was
used to provide a dispersion of 1.5~\AA/pixel and a spectral
resolution of 5.1~\AA\ over the range $\sim4400-5600$~\AA;
the integration time was one hour ($3\times 1200$~s).
Using the data reduction and analysis procedure outlined in Perrett et al.\ (2001),
037--B327 was found to have a heliocentric radial velocity of 
$v = -338 \pm 12$~km~s$^{-1}$ and a metallicity of ${\rm [Fe/H]} = -1.07 \pm 0.20$. 
The spectrum of 037--B327 is shown in Figure~\ref{fig-spectrum},
with the spectrum of another M31 cluster of comparable metallicity
for comparison. The two objects' spectra are very similar, except for
the suppressed blue continuum of 037--B327, and the radial velocity
leaves little doubt that 037--B327 is indeed an M31 globular cluster. 

The photometric measurements summarised in Table~\ref{phot-data}
demonstrate the very red colour of 037--B327.
Barmby et al.\ \shortcite{b00} used their CCD photometry to construct reddening-free 
parameters and estimate the reddening of 037--B327 at $E(B-V)=1.38\pm0.02$. 
This was the largest reliable reddening value in their sample of 221 objects.
Predicting the intrinsic colours from the spectroscopic metallicity
using the method described in Barmby et al.\ \shortcite{b00}, we determine a value
for the reddening of $1.30\pm0.04$, in good agreement with the previous work.
The weighted combination of values from the two methods
gives a final reddening value of $1.32\pm 0.05$.
Assuming $R_V=3.1$ and a distance to M31 of $\mu=24.47$
[Stanek \& Garnavich \shortcite{sg98}; Holland \shortcite{hol98}]
the absolute magnitude of 037--B327 is $M_V=-11.74$. 
The next-brightest
M31 GC, 023--078, has $M_V=-11.36$, while the brightest Milky Way GC is
$\omega$~Cen with $M_V=-10.29$ \cite{h96}. The spectrum used here to
measure 037--B327's radial velocity does not have sufficient resolution
to measure a velocity dispersion, but if 037--B327 has a mass-to-light ratio
typical of M31 GCs [$M/L_V\approx2$; Dubath \& Grillmair \shortcite{dg97}], 
its luminosity implies 
a mass of $8.5\times10^6{\rm M}_{\sun}$. This places it among the most
massive GCs in M31; for example, Meylan et al.\ \shortcite{mey01} find masses in the range
$7-17\times10^6{\rm M}_{\sun}$ for the M31 cluster 000--001.

\begin{table}
\caption{Photometric data for 037--B327\label{phot-data}}
\begin{tabular}{ll}
Measurement & Reference\\
$V=16.92$, $B-V=2.15$ & Kron \& Mayall \shortcite{km60}\\
$V=16.70$, $B-V=2.11$ & Vete\v{s}nik \shortcite{vet62b}\\
$V=16.80$, $B-V=2.17$ & van~den~Bergh \shortcite{vdb69}\\
$V=16.71$, $B-V=2.13$ & Buonanno et al.\ \shortcite{buo82}\\
$V=16.82$, $B-V=2.05$ & Barmby et al.\ \shortcite{b00}\\ 
$K=10.95, J-K=1.25, H-K=0.28$ &B\`{o}noli et~al.\ \shortcite{bo87}\\
$V-R=1.28, V-I=2.63$ &  Barmby et al.\ \shortcite{b00}\\ 
\end{tabular}
\end{table}

\section{Reddening, extinction, and the brightest clusters in M31}

van~den~Bergh \shortcite{vdb68} calculated the absolute magnitudes of the
intrinsically-brightest GCs in M31 by assuming a single intrinsic
colour and using this to determine the colour excess.
He found that if $R_V=3.0$ was adopted, 037--B327 was the brightest M31 GC. 
van~den~Bergh argued that, since 
`there is no {\em a priori} reason why the intrinsically
brightest globular clusters in the Andromeda nebula should
also be the most highly-reddened', the value of $R_V$ in 
M31 was more likely to be 2.5 instead of the Milky Way's 3.0,
as first suggested by Kron \& Mayall \shortcite{km60}.
With this value of $R_V$, he found that 037--B327 became the 
fifth-brightest GC in M31. Table~\ref{brightest} summarises
the photometric data from Barmby et al.\ \shortcite{b00}
and metallicities from Huchra, Brodie \& Kent \shortcite{hbk91}
(for all objects except 037--B327) for the six 
intrinsically-brightest M31 GCs.
For the four objects in common with Table~II of 
van~den~Bergh \shortcite{vdb68}, our values of $E(B-V)$ 
from Barmby et al.\ \shortcite{b00}
are significantly lower. This is because van~den~Bergh's
assumed value of $(B-V)_0=0.6$, corresponding to 
${\rm [Fe/H]}\approx-1.9$, is too low: all of these
clusters have ${\rm [Fe/H]}\geq-1.1$.
These six clusters remain the brightest M31 GCs 
whether $R_V=2.5$ or 3.1 is used, although their 
magnitude ranks change: for $R_V=2.5$, 037--B327 becomes the second-brightest,
not the fifth as van~den~Bergh found. This renders his
argument somewhat less compelling.

\begin{table}
\caption{The Brightest M31 Globular Clusters\label{brightest}}
\begin{tabular}{lclccr}
&&&\multicolumn{2}{c}{Rank in $V_0$}&\\
Name&$V$&$E(B-V)$&$R_V=3.1$&$R_V=2.5$&[Fe/H]\\
000--001&13.75  & 0.08 & 4 & 3 & $-1.08$\\
023--078&14.22  & 0.36 & 2 & 1 & $-0.92$\\
037--B327&16.82 & 1.32 & 1 & 2 & $-1.07$\\
082--144&15.54  & 0.72 & 3 & 4 & $-0.86$\\
151--205&14.83  & 0.38 & 5 & 6 & $-0.75$\\
225--280 &14.15 & 0.15 & 6 & 5 & $-0.70$\\
\end{tabular}
\end{table}

\begin{figure}
\epsfxsize=8cm
\epsfbox{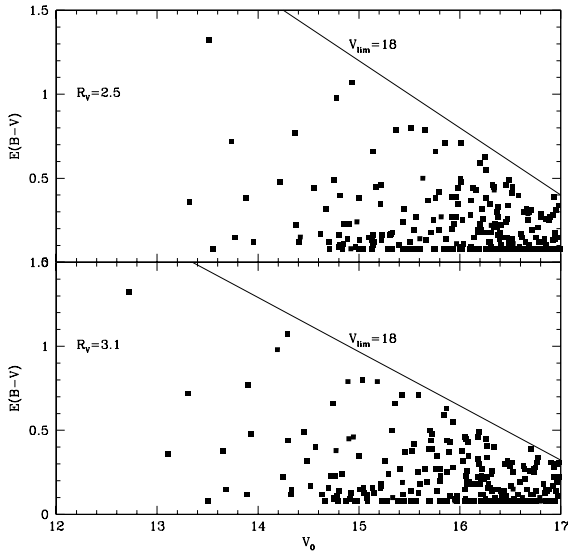}
\caption{$V_0$ vs. $E(B-V)$ for M31 GCs for the $V<18$ sample used by
Barmby et al.\ \shortcite{bhb01}. 
\label{fig-reddist}}
\end{figure}

Four of the six globular clusters in Table~\ref{brightest}
have values for $E(B-V)$ well above the M31 average of 0.22;
the high reddening of the brightest M31 GCs was
part of van~den~Bergh's argument. But is this effect
a persuasive argument for a different $R_V$ in M31?
We believe it is more likely a selection effect.
Assuming that intrinsic luminosity is uncorrelated with observed
extinction, only the intrinsically-brightest clusters in a 
magnitude-limited sample will show the full range of
the extinction distribution. This is shown graphically in 
Figure~\ref{fig-reddist} for the sample of clusters with $V<18$
used by Barmby, Huchra \& Brodie \shortcite{bhb01}. For either $R_V=2.5$ or $R_V=3.1$,
the brighter clusters have a larger range of reddening values.

Examining the entire reddening/magnitude distribution,
we find that the intrinsically brightest 10~per~cent 
of the clusters have an average $E(B-V)$ of 0.42, 0.21~mag larger than
the average value for the other 90~per~cent of clusters.
We modeled this effect by simulating 200 M31 globular cluster
luminosity functions (GCLFs). We drew clusters' extinction-free magnitudes 
from a realistic GCLF with peak $V_0=16.8$ and dispersion ${\sigma}_t=1.0$, 
applying to each an
extinction drawn at random from the observed distribution
for M31 GCs, and computed the observed magnitude. Removing clusters
whose observed magnitudes were fainter than the magnitude limit $V=18$,
we then computed the mean $E(B-V)$ for the brightest 10~per~cent and 
fainter 90~per~cent of the remaining clusters. In more than 90~per~cent of the trials,
the brighter clusters were more reddened, with an average difference
between brighter and fainter clusters of $E(B-V)=0.21$. This is
very similar to what we see in the actual distributions for M31.
The situation where the brightest cluster is also the most reddened
(the case for 037--B327 and $R_V=3.1$) occurs in only 3 of the simulated GCLFs.
However, the situation where the second brightest cluster is
the most reddened (as would be the case for $R_V=2.5$) is no more
common, so we conclude that the case of 037--B327
does not provide sufficient reason to adopt $R_V=2.5$.

The assumption made above that reddening and intrinsic luminosity
are uncorrelated is probably an over-simplification. Figure~8
of Barmby et al.\ \shortcite{b00} shows that the more heavily-reddened clusters are
preferentially located toward the centre of M31, and Barmby et al.\ \shortcite{bhb01}
found that clusters closer to the centre of M31
are, on average, brighter (although this result disagrees with
some previous work on radial variation in the GCLF).
The combination of radial variations in reddening and luminosity
with the magnitude-limit selection effect makes it more likely for
the brightest cluster to be the most heavily-reddened.

Another possible explanation of both the apparent radial GCLF variation 
and 037--B327's unusual properties (both brightest and most reddened)
is the postulate that $R_V$ in M31 increases with distance from
the centre of the galaxy $R_{gc}$. Much of the previous work on this
topic is based on the colour distribution of M31 globular clusters.
Changes in the reddening law are often taken as indicative of
changes in $R_V$; however, the analytic models in Table~3 of
Cardelli, Clayton \& Mathis %
\shortcite{ccm89} indicate that changing $R_V$ from 3.1 to 2.5 
changes $X(UBV)=E(U-B)/E(B-V)$ from 0.72 to 0.75, less than 5~per~cent.
Freedman \& Madore %
\shortcite{fm90} cite unpublished results by L.~Searle showing
a decrease in $X(UBV)$ with distance from the galaxy centre,
but find that either decreasing or constant $R_V$ are
consistent with their M31 Cepheid distance-modulus results.
Iye \& Richter \shortcite{ir85} examined the mean colours and magnitudes of M31 GCs 
in spatial bins, and  measured marginally significant differences 
in $X(UBV)$ between inner and outer clusters. They also 
derived a best-fit value of $R_V=2.6\pm0.7$ for clusters projected
near the major axis, and an average value of $X(UBV)=1.01\pm0.11$.

Using more complete photometric and spectroscopic data,
Barmby et al.\ \shortcite{b00} found that the M31 and Milky Way reddening laws
were the same to within observational error. Examining
those results in more detail, we found no evidence for 
variation in the extinction law with $R_{gc}$.
We attempted to replicate Iye \& Richter's \shortcite{ir85} study
with our more modern data. We found $X(UBV)=0.72\pm0.13$, identical 
to the canonical Milky Way value for $R_V=3.1$, and
no reason to prefer $R_V=2.5$ over $R_V=3.1$ to describe
the distribution of $\langle B-V\rangle$ against $\langle V\rangle$.
While we thus find no evidence for radial variation in the
M31 extinction law from GC colours, studies using other 
methods and classes of objects would be helpful to 
confirm our conclusions.

Aside from being more heavily-reddened, which we have argued
to be a selection effect, the brightest globular clusters in M31 
are not particularly unusual.
As noted above, they tend to be metal-rich; all have ${\rm [Fe/H]}\geq-1.1$,
while the M31 median is ${\rm [Fe/H]}=1.15$ \cite{b00}.
Djorgovski et al.\ \shortcite{dj97} measured velocity dispersions
for 21 clusters including 000--001, 023--078, and 225--280;
these three objects had the three highest velocity dispersions.
All three have values of ${\sigma}_v \sim 25$~km~s$^{-1}$, indicating
that they have the large masses which would be expected for such bright objects.
The cluster 000--001, also known as G1 or Mayall~II, has been 
the subject of several high-resolution imaging studies
\cite{pv84,ric96,mey01}, all of which noted that it was quite
flattened, with $\epsilon=1-b/a \sim 0.2$.
Meylan et al.\ \shortcite{mey01} find that,
like the bright Milky Way cluster $\omega$~Cen, 000--001 shows evidence
for a metallicity spread, and suggest that it
may in fact be the core of a dwarf elliptical galaxy.
Hubble Space Telescope observations of 225--280 \cite{fp96,s01} 
show it to be a very metal-rich (${\rm [Fe/H]}\sim-0.3$, somewhat
higher than the spectroscopic metallicity) but otherwise unremarkable cluster.
The ellipticities of the other brightest clusters, as measured by
Staneva, Spassova \& Golev \shortcite{ssg96}, range from
0.02 to 0.08, well within the range of values for the fainter clusters.
 
\section{Conclusions}

The `remarkable object 037--B327' {\em is} an M31 globular cluster.
It is extremely reddened, with $E(B-V)=1.32\pm0.05$, and
extremely luminous (almost four times as luminous as the brightest Milky Way globular).
However, its radial velocity and metallicity are entirely unremarkable.
We use the M31 globular clusters'
reddening values to examine van~den~Bergh's \shortcite{vdb68}
argument than the value of $R_V$ in M31 differs significantly
from that in the Milky Way. We find that the brighter
clusters are more heavily-reddened, but suggest that this is a combination
of two other effects: selection bias in a magnitude-limited sample,
and variation in the M31 GCLF with distance from the galaxy centre.
A radial variation in $R_V$ in M31 could possibly account for
the GCLF variation and the reddening/magnitude distribution,
but there is at present no evidence for such a variation.

\section*{Acknowledgments}

We thank M.~Irwin for help measuring APM coordinates, J.~Huchra
and D.~Hanes for helpful discussions, and our collaborators on the WYFFOS
project for permission to publish the 037--B327 data 
in advance of the main publication. 
The Digitized Sky Survey was produced at the Space
Telescope Science Institute under U.S. Government grant NAG
W-2166. The images of these surveys are based on
photographic data obtained using the Oschin Schmidt
Telescope on Palomar Mountain and the UK Schmidt Telescope.
The plates were processed into the present compressed
digital form with the permission of these institutions.

\end{document}